# A NEW VALUE OF THE PERIOD OF THE CLASSICAL CEPHEID RT AUR ON THE BASIS OF 456 TIMES OF MAXIMUM, 1897-2023

Guy Boistel
GEOS (Groupe Européen d'Observation Stellaire), http://geos.upv.es/

**Abstract**
The present study collects 456 new times of maximum light of the classical Cepheid RT Aur, covering the period from 1897 to 2022. The O-C diagram resulting from these observations shows that the period given by the GCVS has to be corrected. It results that no strong period variation is found. However, the observed O-C residuals show a long term periodic trend. In the hypothesis of RT Aur being in a binary system, an orbit cannot be deduced from the available astrophysical data.

**Résumé**
Cette étude rassemble 456 nouveaux instants de maximum de lumière de la céphéide classique RT Aur, couvrant la période allant de 1897 à 2022. Le diagramme des O-C résultant de ces observations montre que la période donnée par le GCVS a besoin d'une correction. En conséquence, aucune variation particulière de la période n'est mise en évidence. Cependant, les résidus des O-C observés montrent une variation périodique à long terme. Dans l'hypothèse de l'appartenance de RT Aur à un système binaire, les données astrophysiques actuelles ne permettent pas de déduire une orbite.

## 1. Introduction

**RT Aur** (48 Aur; J2000.0: α = 06h 28m 34.09s ; δ = +30° 29' 34.''; sp. F4Ib-G1Ib) is one of the classical Cepheids for which we have the oldest times of maximum recorded in the scientific literature. The first time of maximum we can process dates back from the year 1897. The variability of 48 Aur has been established by Astbury (1905) and confirmed by Williams (1905) in the same paper (Astbury 1905). RT Aur is a bright star and it is the reason why this star is observed since the end of the 19th century. It is still a favorite among amateurs variable stars observers.

GCVS (Samus et al. 2017) gives the following elements:

- RT Aur varies between V-magnitude 5.00 and 5.82
- Time of maximum ephemeris:
  **HJD = 2454153.880 + 3.728485 × E (1)**

Figure 1 shows the finding chart regularly used by GEOS observers.

Table 1 gives the HD references for the comparison stars.

Table 1: References for the comparison stars.

| A | B | C | D |
|---|---|---|---|
| 44 Aur | 49 Aur | 53 Aur | |
| HD 43039 | HD 46553 | HD 47152 | HD 43646 |

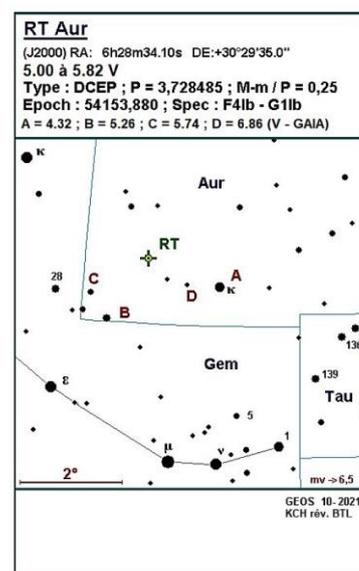

Figure 1: GEOS finding chart for RT Aur. © GEOS – Serge Kuchto.

A huge amount of data is available in the literature and in the various databases provided by automated telescopes or by various amateurs' web sites or servers.

On this wide basis of observations, we are able to study the evolution of its period over more a century. One of the most interesting topics of research about Cepheids is to detect or not a light-time effect (LiTE) in the long term trend in the O-C diagram of these variable stars.

Exploration of the long term behavior of the period of RT Aur has been made by Wunder (1992) (Figure 2a), by Berdnikov (2003) (Figure 2b) and Turner et al. (2007) (Figure 2c) in particular.

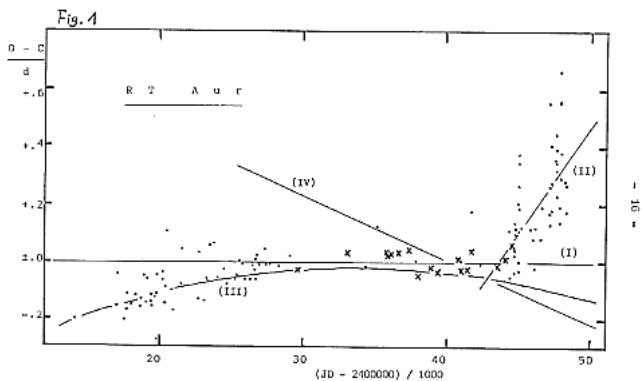 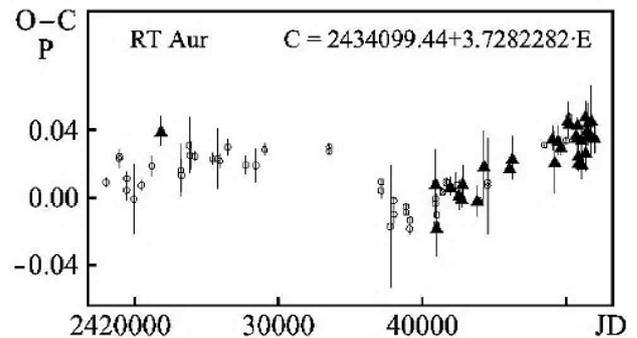

Figure 2a: RT Aur, O-C computed by Wunder (1992).     Figure 2b: RT Aur, O-C computed by Berdnikov (2003).

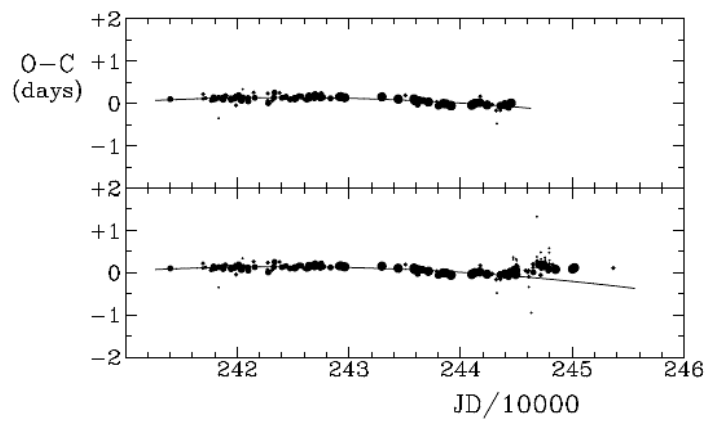

Figure 2c: RT Aur, O-C computed by Turner et al. (2007).

Wunder (1992) and Berdnikov (2003) show the same type of period variation break after JD 2441000, while Turner et al. (2007) suggests a quadratic variation of the period, with fluctuations around the mean value, as we can observe for the classical cepheids SV Vul (Boistel 2022; Csörnyei et al. 2022) or Y Oph (Csörnyei et al. 2022) for example.

As we can see on the figures 2a-2c, the different types of variation highlighted, without being incompatible with each other, do not allow us to definitively decide on the behavior of the period of RT Aur.

With the help of more than 140 of new times of maximum provided by the variable stars observers groups GEOS and BAV, we decided to reinvestigate the long term variation of the period of RT Aur on the basis of this new amount of times of maximum.

## 2. Collecting and processing times of maximum

Figure 3 shows the various sources used to collect times of maximum. Figure 4 shows the different types of photometry merged together for this study.

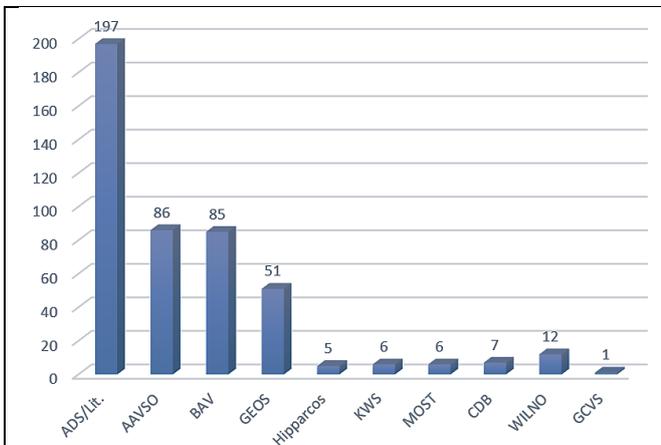

Figure 3: Sources used to collect the observations.

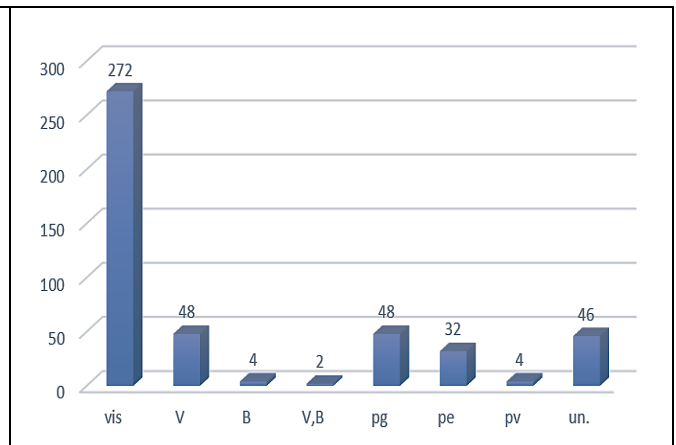

Figure 4: Different types of photometry used for the collected observations.

This study gathers 456 times of maximum found in professional literature (NASA ADS server), computed from various public databases (ASAS-SN, KWS, AAVSO Vstar access) and from unpublished observations (mainly by GEOS members) for this paper as labelled in table 2 (excerpt).

Table 2: Excerpt from the main table of the data used for this study of the behavior of the period of the classical Cepheid RT Aur. This table 2 lists **456** times of maximum of RT Aur.

| ToM +2400000 | Ucrt(d) | Type | E,Phi | P. Shift | E GCVS | O-C (d) | Observer | Reference (see bibliography) | source |
|---|---|---|---|---|---|---|---|---|---|
| 14000.600 | 0.150 | vis | -10769.33 | -1 | -10770 | 2.50 | Müller & Kempf 1899 | Szabados 1977 | ADS |
| 14362.300 | 0.150 | vis | -10672.32 | -1 | -10673 | 2.54 | Astbury&Williams 1905 | Astbury / Hellerich 1937 | ADS |
| 16934.800 | 0.050 | vis | -9982.36 | -1 | -9983 | 2.39 | Astbury&Williams 1905 | Meyer 2023 | ADS |
| 16938.466 | 0.150 | vis | -9981.38 | -1 | -9982 | 2.32 | Astbury&Williams 1905 | Szabados 1977 | ADS |
| 16941.920 | 0.150 | vis | -9980.45 | -1 | -9981 | 2.05 | Astbury&Williams 1905 | BAA Williams, Atsbury, 1905 | ADS |
| 16942.300 | 0.150 | vis | -9980.35 | -1 | -9981 | 2.43 | Williams 1905 | Szabados 1977 | ADS |
| 17173.300 | 0.150 | pg | -9918.39 | -1 | -9919 | 2.26 | Henroteau & Frédette | Henroteau 1925 | ADS |
| 17173.360 | 0.150 | vis | -9918.38 | -1 | -9919 | 2.32 | Williams 1905 | Turner 2007 tab.2 | ADS |
| 17173.459 | 0.150 | vis | -9918.35 | -1 | -9919 | 2.42 | Williams | Williams PA 1916 | ADS |
| 17639.320 | 0.050 | vis | -9793.40 | -1 | -9794 | 2.22 | Von Zeipel | Meyer 2023 | ADS |
| 17643.048 | 0.150 | vis | -9792.40 | -1 | -9793 | 2.22 | Zeipel 1908 | Szabados 1977 | ADS |
| 17781.070 | 0.050 | pg | -9755.39 | -1 | -9756 | 2.29 | Wendell 1913 | Meyer 2023 | ADS |
| 17788.100 | 0.150 | vis | -9753.50 | 0 | -9754 | 1.86 | Wendell 1913 | Szabados 1977 | ADS |
| 17788.484 | 0.150 | vis | -9753.40 | -1 | -9754 | 2.25 | Wendell 1913 | Szabados 1977 | ADS |

To be continued… (see below for the links to download the data).

Columns are set as follows:
- Col.1: the time of maximum (ToM after) (JD +2400000.0);
- Col.2: the uncertainty on the ToM in days if known (un. for "unknown");
- Col.3: the type of photometry if known;
- Col.4: the value of (cycle, phase) computed on ephemeris (1);
- Col.5: the period shift (see below);
- Col.6: the value of cycle E computed on ephemeris (1), taking in account the period shift;
- Col.7: the O-C in day computed on the GCVS ephemeris (1);

- Col.8: the observer if known (un. for "unknown");
- Col.9: the bibliographical reference (see the bibliography)
- Col.10: the code of the source (ADS for literature; WILNO for Wilno observatory in Poland; AAVSO; BAV; CDB for McMaster Cepheid Database; GEOS; Hip for Hipparcos; KWS; MOST for satellite MOST).

The data can be downloaded from the GEOS Web site, publications data page:
- Excel .xlsx format
- .csv format
- .txt format (with tabulations)

In more details, we have collected times of maximum from AAVSO, BAV and GEOS databases and publications. VSTAR software is a useful tool to access to the visual and CCD observations performed by AAVSO members. We have processed a part of visual AAVSO data with the help of the softwares Peranso© and Mavka© for time of maximum determination.
BAV data can be downloaded from the BAV web site and the option "Data for scientists". Lienhard Pagel, Chairman of the BAV, provided Ralf Meyer's collection of Cepheids times of maximum and O-C diagrams (Meyer 2023a, 2023b and 2023c).
RT Aur is one of the Cepheids observed from the beginnings of the GEOS in the 1970's. An old GEOS collection of times of maximum has been published by Boistel (2023) in GEOS open access publications. The bibliography indicates the other GEOS publications relative to RT Aur (Boninsegna 1982, Busquets 1982, Colombo 1989, Dalmazzio 1997, Dalmazzio 1999, Gobet 1989, Gobet 1992).

We used the CCD V observations provided by the automated telescope KWS accessible on their original servers and via the AAVSO-VSX server as well. We made also use of the original data recorded on the McMaster Cepheid International Database (CDB) (Berdnikov 1995). As we can see on figure 2, the AAVSO, BAV and GEOS groups provide a great part of the O-C collected here and it is not surprising that close to 60% of the times of maximum collected come from visual observations.

### 3. Analysis of the data: the period

On the basis of the 456 times of maximum collected given in table 2 (excerpt), it is possible to draw the O-C diagram of RT Aur. Figure 5 shows the O-C diagram versus the cycle number computed on GCVS ephemeris (1) for the various sources of observations. It has been necessary to shift the cycle numbers for the times of maximum from cycles -11000 to -8000 as indicated in excerpts of the table 1 to avoid wrong interpretation of the change of the period. In fact we can see that the amplitude of the O-C variation (3 days) is higher than the variation taken in account in the main published studies given O-C variations for RT Aur (Berdnikov et al. 2003, Fernie 1993, Meyer 2023c, Turner et al. 2007, Szabados 1991). These studies used an incomplete sample of available data and mainly focused on small amplitude variations around the cycle of number -2000 (for the ephemeris (1)).

From figure 5, the clear long term linear trend shows that the ephemeris and the period has just to be corrected. The interpretation of this diagram of O-C is then quite simple. We need only to process a linear fit to obtain the new ephemeris (2) below.

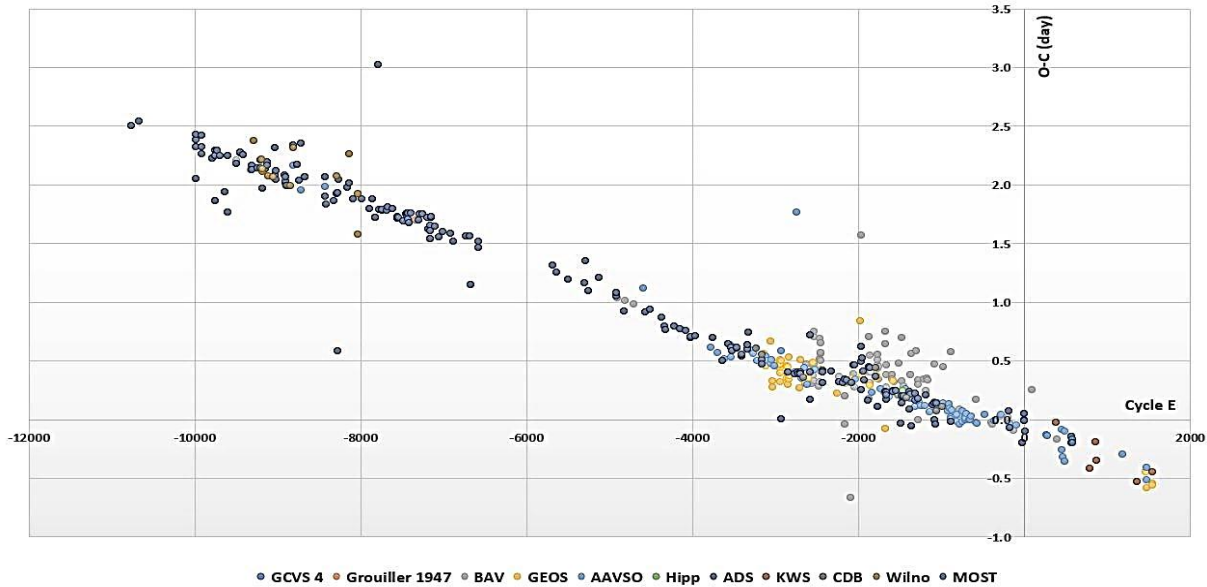

● GCV5 4  ● Grouiller 1947  ● BAV  ● GEOS  ● AAVSO  ● Hipp  ● ADS  ● KWS  ● CDB  ● Wilno  ● MOST

Figure 5: RT Aur, diagram of O-C computed on ephemeris (1), for the various sources of observations. On the x-axis: the cycle E ; on the y-axis, O-C in days. The linear trend is obvious.

The linear ephemeris obtained by a linear fit is the following:
**MAX: HJD = 2454153.743   +   3.728 2403 × E (2)**
$$\pm \qquad .013 \quad \pm \quad .000\ 0028$$

Figure 6 shows the linear residuals computed with ephemeris (2). The mean of the residual is equal to $1.6 \times 10^{-11}$ day we can consider negligible. The standard deviation is: $\sigma = 0.188$ day. Several times of maximum are far from the mean of the times of maximum collected and closed to the usual limit of $3 \times \sigma$. But we have decided to not exclude these times of maximum for further investigations.

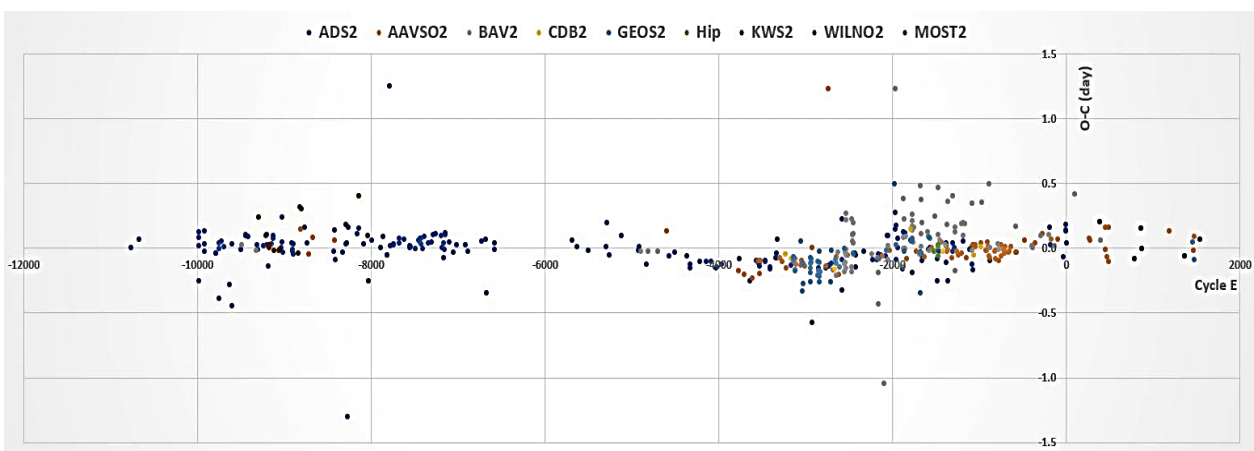

Figure 6: Linear O-C residuals for RT Aur computed on ephemeris (2). A slow sinusoidal trend is visible on this diagram. On the x-axis: the cycle E ; on the y-axis: O-C residuals of the linear fit.

### 4.  Analysis of the data: the long term sinusoidal trend

Figure 6 shows clearly a slow sinusoidal trend as an indicator of a possible existence of a companion of the variable star. Evans (1983), Szabados (1988), Turner et al. (2007) and Evans et al. (2015) for example, give good indicators for the explanation of the periodic fluctuations detected in the O-C residuals diagrams of some Cepheids, FN Aql, RX Aur, AW Per for example, and RT Aur.

Evans et al. (2015) provides an account of the several causes suggested to explain period variations in Cepheids, both fundamental mode pulsators and first overtone pulsators:

(a) *Evolution through the Cepheids instability strip*: one direction of period change predominates and it results in a parabolic O-C diagram.

(b) *Light-time effects in binary systems:* this effect produces cyclic apparent period changes. But this must be consistent with the elements of the orbit of the system.

(c) *Star spots*: this cause has been suggested for the variations of the Kepler Cepheid V1154 Cyg. But while starspots could affect the time of maximum light, they would not have a cumulative effect as seen in the O-C diagram.

(d) *Mass-loss*: increasing periods and quadratic variations of the O-C observed in some Cepheids can be partially explained by mass-loss cause but it is only one of the several factors of the variations observed.

(e) *Pulsation and Blazhko effect*: if the phenomenon is not fully understood, pulsation and convection may drive pulsation mode excitation and hence amplitude variation, and the same might also affect Cepheids periods.

The mean amplitude of the residuals obtained from figure 6 is about 0.24 jours, which is larger than the amplitudes discussed in the quoted papers above and already considered too high.

Turner et al. (2007) suggested a possible light-time effect to explain the O-C residuals for RT Aur. But Evans et al. (2015) shows that the available velocity variations measurements indicate no variation larger than $\pm 1$ km.s$^{-1}$ ; this means that no orbital velocity variation has been observed for data obtained over a long time interval. But let's notice that only one third of the orbital cycle is covered observationally (Turner et al. 2007). Evans (1983) noticed that for most Cepheids, radial velocities were ever lacking, as we can see in the data collected by Berdnikov (1995) for the McMaster Cepheid archives.

This means that the fluctuations observed for RT Aur cannot be explained by only one of the factors (a) to (e) listed above.

## 5. Conclusion

Our study reinvestigates the period change of the classical Cepheid RT Aur, a fundamental pulsator. On the basis of a huge collection of new times of maximum light, the period has been corrected and the long term trend of its variation is shown purely linear.

A new value of the period is established as follows:

**MAX: HJD = 2454153.743  +  3.728 2403 × E (2)**

**                    ±       .013   ±    .000 0028**

From the O-C residuals, sinusoidal periodic variations are highlighted. The available scientific literature underlines the lack of data concerning the orbital follow-up of this star, in order to clearly establish the causes of the observed fluctuations in the O-C residuals and in the period change of RT Aur.

Most authors call for a spectrophotometric monitoring of RT Aur to clarify variations in its period and to obtain a fully orbital solution.

## Bibliography and references

BAV Journal search: https://bav-astro.de/index.php/veroeffentlichungen/bav-online-journal/journalsuche

BAV Rundbrief search: https://www.bav-astro.eu/Rundbriefe/

OEJV issues: http://var.astro.cz/oejv/oejv.php?lang=en.

GCVS (Samus et al. 2017)

GEOS website: http://geos.upv.es/ and unpublished observations performed by Michel Dumont, Guy Boistel, Stéphane Ferrand (private communications)

GEOS Open Access Publications: http://geos.upv.es/index.php/publications/NCOA/

KWS (Kamogata/Kiso/Kyoto Wide-field Survey): http://kws.cetus-net.org/

McMaster Cepheid Photometry and Radial Velocity Data Archive: https://physics.mcmaster.ca/Cepheid/ (Berdnikov 1995)

McMaster data archives - Data for RT aur: International photometry database of Cepheids – RT Aur

PERANSO software: https://www.cbabelgium.com/peranso/index.html (Paunzen & Vanmunster 2016)